\documentclass[sigconf,natbib=true]{acmart}
\AtBeginDocument{%
  \providecommand\BibTeX{{%
    \normalfont B\kern-0.5em{\scshape i\kern-0.25em b}\kern-0.8em\TeX}}}

\usepackage{enumitem}
\usepackage{multirow}
\usepackage{adjustbox}
\usepackage{pifont}
\usepackage{xcolor}
\usepackage{arydshln}
\usepackage{cancel}
\newcommand{\cmark}{\color{green!70!black}\ding{51}}%
\newcommand{\xmark}{\color{red}\ding{55}}%

\copyrightyear{2024} 
\acmYear{2024} 
\setcopyright{rightsretained} 
\acmConference[SIGIR '24]{Proceedings of the 47th International ACM SIGIR Conference on Research and Development in Information Retrieval}{July 14--18, 2024}{Washington, DC, USA}
\acmBooktitle{Proceedings of the 47th International ACM SIGIR Conference on Research and Development in Information Retrieval (SIGIR '24), July 14--18, 2024, Washington, DC, USA}
\acmDOI{10.1145/3626772.3657843}
\acmISBN{979-8-4007-0431-4/24/07}

\begin{document}

\title{Towards Human-centered Proactive Conversational Agents}


\author{Yang Deng}
\affiliation{\institution{National University of Singapore}\country{Singapore}}
\email{ydeng@nus.edu.sg}

\author{Lizi Liao}
\affiliation{\institution{Singapore Management University}\country{Singapore}}
\email{lzliao@smu.edu.sg}

\author{Zhonghua Zheng}
\affiliation{\institution{Harbin Institute of Technology, Shenzhen}\country{China}}
\email{polang1999@gmail.com}

\author{Grace Hui Yang}
\affiliation{\institution{Georgetown University}\country{United States}}
\email{grace.yang@georgetown.edu}

\author{Tat-Seng Chua}
\affiliation{\institution{National University of Singapore}\country{Singapore}}
\email{chuats@comp.nus.edu.sg}


\begin{abstract}

Recent research on proactive conversational agents (PCAs) mainly focuses on improving the system's capabilities in anticipating and planning action sequences to accomplish tasks and achieve goals before users articulate their requests. This perspectives paper highlights the importance of moving towards building human-centered PCAs that emphasize human needs and expectations, and that considers ethical and social implications of these agents, rather than solely focusing on technological capabilities. The distinction between a proactive and a reactive system lies in the proactive system's initiative-taking nature. Without thoughtful design, proactive systems risk being perceived as intrusive by human users. We address the issue 
by establishing a new taxonomy concerning three key dimensions of human-centered PCAs, namely \textbf{\textsc{Intelligence}}, \textbf{\textsc{Adaptivity}}, and \textbf{\textsc{Civility}}. 
We discuss potential research opportunities and challenges based on this new taxonomy upon the five stages of PCA system construction. 
This perspectives paper lays a foundation for the emerging area of conversational information retrieval research and paves the way towards advancing human-centered proactive conversational systems. 
\end{abstract}





\maketitle

\section{Introduction}

With the advent of large language models (LLMs), the emergence and integration of conversational systems mark a significant leap forward in information retrieval (IR), which evolves many traditional interactive IR systems into conversational IR systems. 
For instance, Microsoft recently released a new version of Bing with its integration with ChatGPT \cite{chatgpt} under the idea of conversational search. 
In the rapidly evolving field of conversational systems, proactive conversational agents (PCAs) \cite{sigirap23-tutorial,sigir23-tutorial,cikm22-pasir} are emerging to revolutionize how systems interact with human users. In the literature \cite{ijcai23-survey,ijhci21-chatbot}, the proactivity of a conversational system typically refers to the system's ability of being aware of the long-term conversational goal and capable of taking initiatives to lead the conversation towards the goal. 
Recent years have witnessed a number of advanced designs that address proactivity on a range of conversational systems. For instance, in conversational information seeking, PCAs are developed to further eliminate the uncertainty for more efficient and precise information seeking by initiating ambiguity clarification~\cite{sigir19-clariq,www20-clariq} or eliciting the user preference \cite{cikm18-crs,wsdm20-ear}, instead of simply reacting to user queries.  
While in open-domain dialogue systems, different from passively echoing the user-initiated discussion topics or emotion requirements, various PCA designs arise to be capable of directing the conversations \cite{sigir22-proactive,tois23-mgcrs}.  
The distinction between a proactive and a reactive system lies in the proactive system's initiative-taking nature, effectively increasing the number of initiators in an interactive system from one (the user) to two (both the user and the machine). Without thoughtful design, proactive systems risk being perceived as intrusive by human users. Consequently, the key to the widespread acceptance and effectiveness of PCAs lies in their design being fundamentally human-centered, rather than solely advancing technical efficiency and proficiency.

\begin{figure}[t]
\setlength{\abovecaptionskip}{2pt}   
\setlength{\belowcaptionskip}{0pt}          
\centering      
\includegraphics[width=0.48\textwidth]{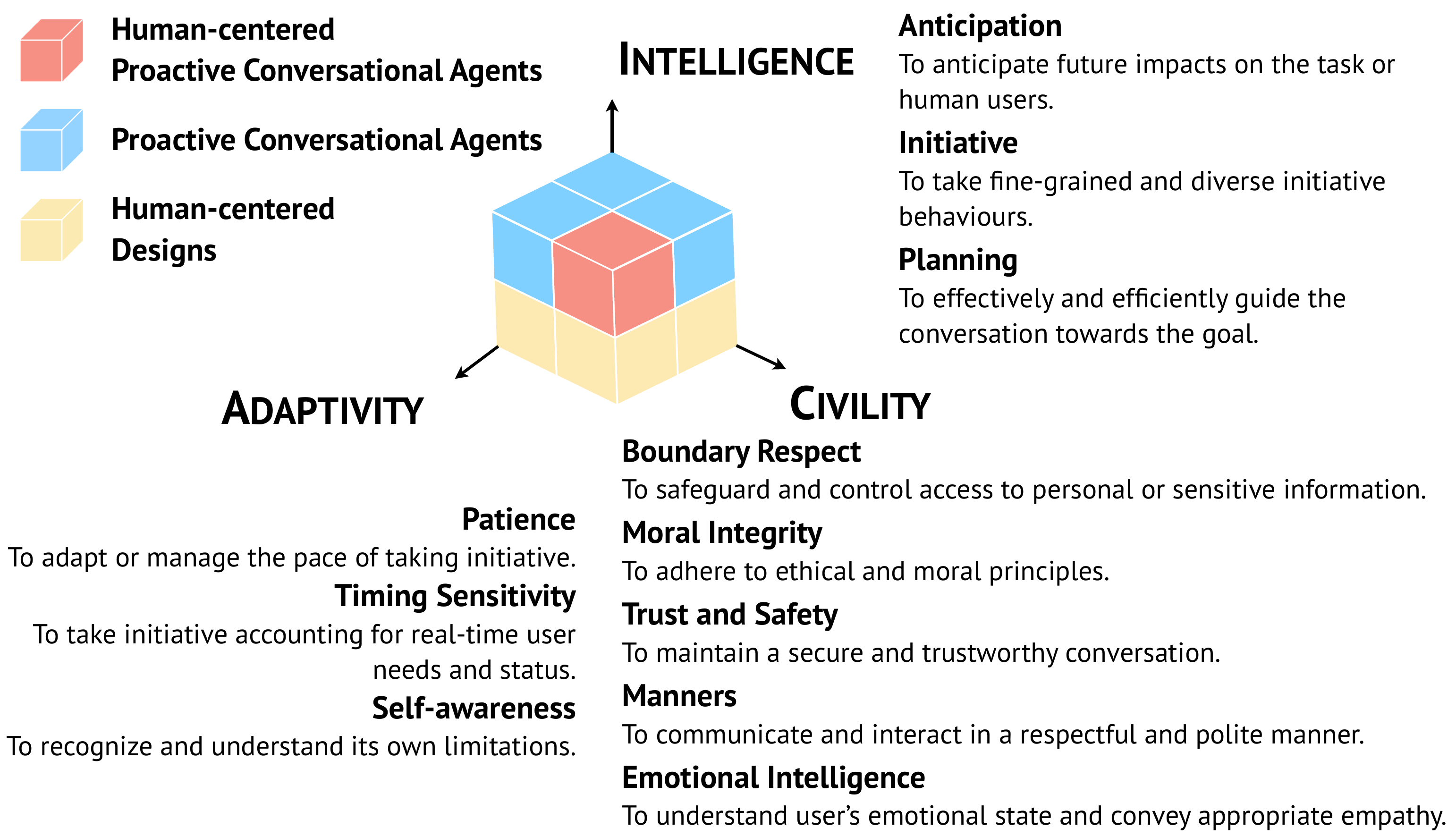}
\caption{Three key dimensions of human-centered proactive conversational agents with representative abilities.}
\label{fig:perspective}
\vspace{-3mm}
\end{figure}

To this end, this perspectives paper discusses the intricate balance of technological advancement and human-centered design principles in the creation of proactive conversational agents. We envision human-centered PCAs to be a kind of PCA that \textit{emphasizes human needs and expectations}, and \textit{considers the ethical and social implications} of these agents, beyond technological capabilities. We propose to establish a new taxonomy concerning three key dimensions of human-centered PCAs, namely \textbf{\textsc{Intelligence}}, \textbf{\textsc{Adaptivity}}, and \textbf{\textsc{Civility}}, as shown in Figure \ref{fig:perspective}. 
We first investigate the past work on proactive conversational agents based on the new taxonomy and then prospect a board research agenda for building human-centered PCAs. 
The goal of this paper is to act as a handbook for discussions on the human-centered designs in every stage of the construction of PCAs, including Task Formulation, Data Preparation, Model Learning, Evaluation, and System Deployment.

\section{Overview}\label{sec:overview}

\subsection{Key Dimensions}

To develop human-centered proactive conversational agents, we have identified three key dimensions specific to these agents. We propose deriving both design principles and construction guidelines from these dimensions to inform the development of human-centered proactive conversational systems. These dimensions are \textbf{\textsc{Intelligence}}, \textbf{\textsc{Adaptivity}}, and \textbf{\textsc{Civility}} (shown in Figure \ref{fig:perspective}). 
\vspace{-0.1cm}
\begin{itemize}[leftmargin=*]
    \item \textbf{\textsc{Intelligence}}: Intelligence in a proactive conversational agent is characterized by its capabilities to anticipate future development of the task and to perform strategic planning ahead of user requests, essential for achieving the conversation's goals proactively. This involves taking nuanced initiative and anticipating both the short-term and long-term impacts on the task or human users. PCAs with low-level intelligence may exhibit inaccurate, and unfocused initiative, like a well-intentioned but amateurish helper, eager to assist but lacking expertise or skills. 
    \item \textbf{\textsc{Adaptivity}}: 
    Adaptivity refers to the capability of PCAs to dynamically adjust and manage the timing and pacing of its actions and interventions in response to the user's real-time context and evolving needs. This requires the agent to be designed with patience in determining the initiative's pace, sensitivity to the impact of taking initiative while considering real-time user needs and status, and self-awareness of its capabilities and limitations, particularly in understanding when and how often to intervene in a manner that is most beneficial and relevant to the user. 
    \item \textbf{\textsc{Civility}}: 
    Civility in proactive conversational agents refers to the agent's capability to recognize and respect the physical, mental, and social boundaries set by the user, the conversational task, and general ethical standards. These agents should be adept at understanding both personal and task-related boundaries and respecting them while taking proactive initiatives. This covers a broad spectrum of personal and social norms, including maintaining privacy, ensuring trust and integrity, and avoiding interactions that are intrusive or disrespectful. 
\end{itemize}
\vspace{-0.1cm}

\subsection{Types of Proactive Conversational Agents}
\begin{figure}[t]
\setlength{\abovecaptionskip}{2pt}   
\centering      
\includegraphics[width=0.34\textwidth]{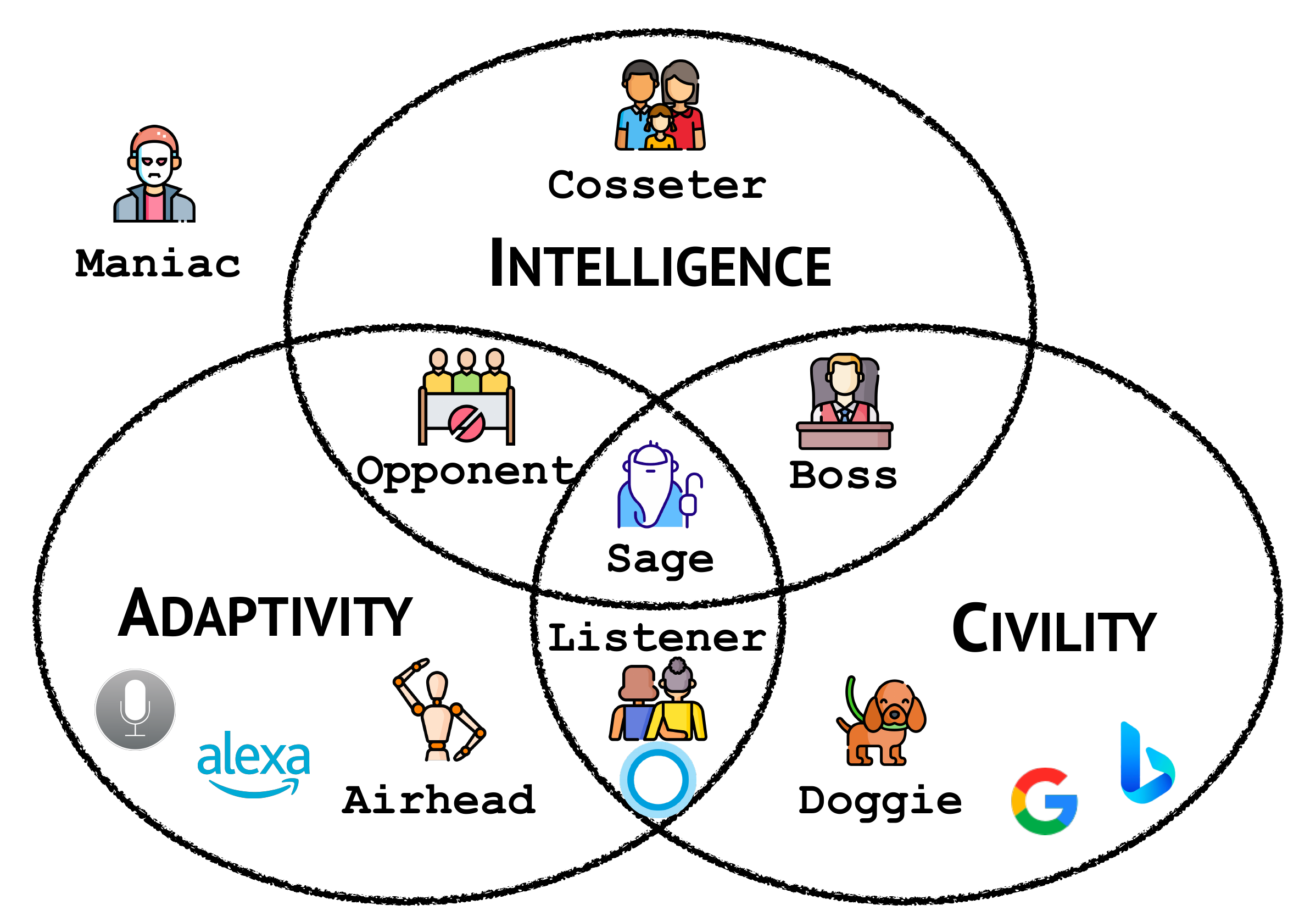}
\caption{Different types of proactive conversational agents in terms of three key dimensions of human-centered PCAs.}
\label{fig:category}
\vspace{-0.45cm}
\end{figure}
\begin{table*}[]
\setlength{\abovecaptionskip}{2pt}   
\setlength{\belowcaptionskip}{2pt}
    \centering
    \caption{Case studies of task formulation. $\text{\cmark}$ and $\text{\xmark}$ denote whether certain dimension has been considered or not.}
    \begin{adjustbox}{max width=1.0\textwidth}
    \begin{tabular}{llllr}
    \toprule
    \textbf{Task Formulation} & \textbf{\textsc{Intelligence}} & \textbf{\textsc{Adaptivity}} & \textbf{\textsc{Civility}} & \textbf{PCA Type}\\
    \midrule
    Asking Clarifying Question & {\xmark ~ (single-type strategy)} & {\xmark ~ (frequently take initiative)} & {\cmark} & \texttt{Doggie} \\
    Mixed-initiative Information Seeking & {\cmark (multi-type strategy)} &{\cmark (depend on initiative needs)} &{\cmark} & \texttt{Sage}\\
    \hdashline
    Empathetic Dialogue & {\xmark ~ (single-type strategy)} & {\cmark} & {\cmark} & \texttt{Listener} \\
    Emotional Support Dialogue &{\cmark (multi-type strategy)} &{\cmark}  &{\cmark} & \texttt{Sage}\\
    \hdashline
    Negotiation Dialogue & {\cmark} & {\cmark} & {\xmark ~ (no restriction on strategies)} & \texttt{Opponent} \\
        Pro-social Negotiation Dialogue & {\cmark} & {\cmark} & {\cmark ~ (constrained by social norms)} & \texttt{Sage} \\
        \hdashline
    Target-guided Dialogue & {\cmark} & {\xmark ~ (favour aggressiveness)} & {\xmark ~ (no restriction on targets)} & \texttt{Cosseter}\\
    Personalized Target-guided Dialogue & {\cmark} & {\cmark ~ (considering user engagement)} & {\cmark ~ (constrained by user preference)} & \texttt{Sage}\\
    \bottomrule
    \end{tabular}
    \end{adjustbox}
    \label{tab:type}
    \vspace{-3mm}
\end{table*}

As illustrated in Figure \ref{fig:category}, based on the proficiency level of the three dimensions, we can categorize the proactive conversational agents into eight general types: 
\begin{itemize}[leftmargin=*] 
    \item \textbf{\texttt{Sage}} (High \textsc{Intelligence}, High \textsc{Adaptivity}, High \textsc{Civility}) denotes the type of PCAs that meet the standards of three dimensions of human-centered PCAs. Its sophisticated, personalized, and respectful interactions, making it a valuable asset in diverse fields that require nuanced and human-centered AI assistance. 
    \item \textbf{\texttt{Opponent}} (High \textsc{Intelligence}, High \textsc{Adaptivity},  Low \textsc{Civility}) denotes the type of PCAs that are designed to engage in thorough and persistent interactions, but possibly challenging or negotiating the user's views and decisions in an opposite position. In order to achieve their specific goals, these systems sometimes may intrude the user's personal or social boundaries. 
    For example, negotiation systems sometimes involve strategies \cite{eacl21-resist-strategy} that can be intrusive and potentially disrespectful to human users, such as attacking the opponent's stance.
    \item \textbf{\texttt{Boss}} (High \textsc{Intelligence}, Low \textsc{Adaptivity},  High \textsc{Civility}) denotes the type of PCAs that would likely offer efficient assistance and be considerate of the user's boundary and privacy, but their interactions might be more direct and to-the-point, prioritizing effectiveness and clarity over user needs and engagement, just like the authoritative boss at work.  
    The analysis of proactive robotic assistants in HCI studies \cite{umap20-timing,icmi21-timing} reveals that taking initiative at different timing leads to different impacts on the user's trust. In high-stakes, fast-paced settings like emergency response or critical business decisions, a \texttt{Boss}-type PCA excels by providing clear, direct guidance while respecting boundaries, ensuring swift and accurate task completion. 
    \item \textbf{\texttt{Cosseter}} (High \textsc{Intelligence},  Low \textsc{Adaptivity},  Low \textsc{Civility}) denotes the type of PCAs that are overly involved and excessively monitoring or controlling in their interactions with users, similar to the overprotective behaviors associated with helicopter parenting. For instance, some conversational recommender systems may excessively acquire users' personal information by asking preference elicitation questions  \cite{sigir21-crs}, which potentially lead to user discomfort or a sense of intrusion, as it may be perceived as invasive or aggressive in its attempt to reach their goals. 
    \item \textbf{\texttt{Listener}} (Low \textsc{Intelligence},  High \textsc{Adaptivity},  High \textsc{Civility}) denotes the type of PCAs that are friendly, engaging, and empathetic in their interactions. These systems are often referred to social chatbots, such as Cortana, which would likely be designed to entertain users by shifting topics \cite{emnlp21-topic-shift} or comforting users. 
    Generally speaking, in casual, low-stress environments such as home settings or social spaces, a \texttt{Listener}-type PCA thrives by offering companionship and emotional support, engaging users with empathy \cite{acl19-empathetic} and flexible conversation. 
    \item \textbf{\texttt{Airhead}} (Low \textsc{Intelligence}, High \textsc{Adaptivity}, Low \textsc{Civility}) denotes the type of PCAs that are perceived as lacking depth, sophistication, or serious functionality, similar to the colloquial use of "airhead" to describe someone who is not very thoughtful or intelligent. For example, the proactivity in voice assistants (\textit{e.g.}, early versions of Siri and Alexa) lies on agent-initiated interactions triggered by contextual and environmental events or user behaviours \cite{proactive-voice-assistant}. Their simplistic yet responsive nature makes them suitable for straightforward tasks or as novice-friendly, unintimidating interfaces for technology newcomers. While users raise concerns on privacy protection and intrusiveness \cite{cui21-voice-assistant,cui22-voice-assistant}. 
    \item \textbf{\texttt{Doggie}} (Low \textsc{Intelligence}, Low \textsc{Adaptivity}, High \textsc{Civility}) denotes the type of PCAs that are friendly, highly responsive, and possibly intuitive, like the typical characteristics associated with dogs. For example, recent years witnessed that many search engines, such as Google and Bing, are equipped with conversational features for proactive interactions, such as suggesting useful queries \cite{www20-suggest-query} or asking clarification questions \cite{sigir19-clariq,www20-clariq}. A user study in \citet{ipm23-user-clariq} shows that systems should ask clarification questions only when necessary, instead of frequently asking clarification questions or suggesting useful queries. 
    \item \textbf{\texttt{Maniac}} (Low \textsc{Intelligence}, Low \textsc{Adaptivity}, Low \textsc{Civility}) denotes the type of PCAs characterized by their aggressive and irrational initiative behaviours, closely resembling the unpredictable nature of an uncontrollable maniac.  
\end{itemize}
\vspace{-0.1cm}

The current landscape of commercial conversational systems predominantly features a foundational level of proactive \textsc{Intelligence}, highlighting an exciting area for ongoing research to elevate their capabilities in this domain. Additionally, while significant strides have been made in developing PCAs, there is an emerging recognition of the importance of further exploring two other vital dimensions: \textsc{Adaptivity} and \textsc{Civility}. These aspects are essential for crafting PCAs that are truly centered around human needs and preferences, offering a well-rounded and user-friendly experience.

\vspace{-0.1cm}

\subsection{Five Stages for PCA System Construction}
In terms of PCA system construction, we can briefly organize the process into five stages sequentially, namely Task Formulation, Data Preparation, Model Learning, Evaluation, and System Deployment.
The proposed human-centered designs are supposed to be present in every stage during the construction of PCAs: 
\begin{itemize}[leftmargin=*]
    \item \textbf{Task Formulation} is the initial stage where the objectives and scope of the PCA are defined, setting its development foundation.
    \item \textbf{Data Preparation} involves collecting, cleaning, and organizing the necessary data that is used for the PCA to learn.
    \item \textbf{Model Learning} is the phase where the PCA is trained using algorithms and the data to make desired decisions and responses.
    \item \textbf{Evaluation} is a critical stage where the agent's performance is assessed to ensure it meets the desired standards of interaction. 
    \item \textbf{System Deployment} is the final stage where the developed PCA is integrated into the environment to interact with users.
\end{itemize}

In what follows, we first re-interpret the current studies of building PCAs from the new human-centered taxonomy upon the five stages for PCA construction, and then correspondingly prospect future research directions and challenges under each stage.

\section{Task Formulation}\label{sec:task}
Existing task formulation of PCAs mainly prioritizes the dimension of \textsc{Intelligence} aimed at goal completion, while it is also crucial to ensure the integration of user emotions, preferences, and values, and adherence to ethical standards. The other two dimensions, \textsc{Adaptivity} and \textsc{Civility}, are key to fostering trust, satisfaction, and seamless interactions between humans and systems. 

With the three key dimensions from human-centered perspectives, we re-interpret the task formulation of existing PCA literature, and discuss how new tasks can be derived from them. As shown in Table \ref{tab:type}, we elaborate the discussions with several widely-studied and representative research task formulations in IR community. 
\begin{itemize}[leftmargin=*]
\item  \textbf{Current: Asking Clarifying Question}. A proactive conversational information-seeking system \cite{ftir-cis,sigir22-cis} might ask for clarification on an ambiguous query. However, frequent clarification requests can negatively impact user experience \cite{tois23-clariq}, resembling a \texttt{Doggie}-type PCA's approach. 
\item \textbf{Desired: Mixed-initiative Information Seeking}. 
Key system-initiated behaviors include asking clarifying questions \cite{sigir19-clariq}, managing out-of-scope queries \cite{tacl23-inscit}, and providing extra information \cite{naacl22-ketod}. 
Recent studies \cite{tacl23-inscit, ijcai23-tod} focus on diverse strategies to enhance the agent's \textsc{Intelligence}. Besides, system initiative prediction \cite{cikm23-system-init} (\textit{e.g.}, clarification need prediction \cite{emnlp21-clariq, pacific}) is vital for improving \textsc{Adaptivity} in mixed-initiative information seeking.
\vspace{0.05in}
\item \textbf{Current: Empathetic Dialogues}. Traditional task formulation \cite{aaai18-emo-chat,acl19-empathetic} on emotion-aware dialogue systems has predominantly focused on crafting responses that echo the user’s emotions or mirror their feelings, being a \texttt{Lisenter}-type PCA. 
\item \textbf{Desired: Emotional Support Dialogues}. In contrast, emotional support dialogue systems \cite{esconv} are designed with the objective of improving the user’s emotional well-being from certain negative emotional states with interventions like in Cognitive Behavioral Therapy (CBT). The task  is formulated to extend beyond merely demonstrating empathy; they should proactively take different emotional support strategies to engage with the user's concerns and deliver actionable advice or encouragement to aid in resolving the issues. 
An empirical analysis \cite{acl23-kemi} shows that proactive behaviours at different phases of the conversation may lead to different impacts on the user’s emotional state. 
Similarly, it is also a crucial subtask for emotional support dialogues to determine when to take the initiative, ensuring \textsc{Adaptivity}. 
\vspace{0.06in}
\item \textbf{Current: Negotiation Dialogues}. 
Negotiation dialogue \cite{negotiate-survey} is a process of strategic interaction aimed at finding mutually acceptable solutions between parties, but, meanwhile, maximizing the profit from one side. This concept, deeply rooted in psychology, political science, and communication, has a wide range of applications in everyday life, including price bargaining, strategic gaming, and persuasion. To successfully negotiate, \textsc{Intelligence} and \textsc{Adaptivity} are key components considered in the current task formulation. This indicates that the prevalent formulation typically focuses on building \texttt{Opponent}-type PCAs for negotiation dialogues, neglecting the perspective of \textsc{Civility}. Strategy modeling is the primary aspect formulated in negotiation dialogue problems. Negotiation strategies encompass various tactics aimed at achieving goals, but some can be intrusive and potentially disrespectful. Real-world negotiation sometimes involves strategies \cite{eacl21-resist-strategy} like contesting (attacking the opponent's stance), empowerment (emphasizing personal preferences to counter claims), and self-pity (evoking guilt). When used inappropriately, these tactics can cross the trust boundary of human users, harming the ethical conduct of the negotiation.  
\item \textbf{Desired: Pro-social Negotiation Dialogues}. To maintain boundary respect, the task formulation should involve constraints of avoiding strategies that might humiliate or provoke the other party and promoting polite and empathetic interactions \cite{naacl22-empathetic-persuasion,coling22-polite}. 
\vspace{0.065in}
\item \textbf{Current: Target-guided Dialogues}. This task involves conversational agents proactively leading the dialogue towards a specific target, \textit{e.g.,} certain topics for chit-chats \cite{acl19-tgc} or particular items for recommendation \cite{acl20-durecdial}. This approach has gained significant attention for its potential to enhance system effectiveness. However, current formulations often neglect \textsc{Adaptivity} and \textsc{Civility}, aligning with a \texttt{Cosseter}-type PCA. In terms of \textsc{Adaptivity}, LLM-based systems \cite{emnlp23-procot} show proficiency in goal-directed conversation, but abrupt topic shifts can reduce user satisfaction and engagement \cite{sigir22-proactive}, resembling aggressive sales tactics. Regarding \textsc{Civility}, current task formulations do not impose restrictions on the choice of targets. If a target topic is harmful or toxic, the conversation may violate ethical boundaries. Similarly, if a target item is chosen solely by the seller without considering user preferences, it can erode users' trust, as they may feel the system prioritizes profits over their needs. 
\item \textbf{Desired: Personalized Target-guided Dialogues}. When considering \textsc{Adaptivity} in target-guided dialogues, a human-centered task formulation should encompass not only the efficiency  of achieving the target but also the constraints related to user satisfaction and the smoothness of topic transitions. Besides, the target should align with the user's interests and needs. For example, in target-guided conversational recommendation \cite{acl20-durecdial,emnlp23-promotion}, the target can be first customized from a set of items for promotion based on user preferences, instead of being directly assigned. 
\end{itemize}

\begin{table*}[]
\setlength{\abovecaptionskip}{2pt}   
\setlength{\belowcaptionskip}{2pt}
    \centering
    \caption{Analysis of user needs and toxicity in existing proactive conversation datasets. (The lower the better $\downarrow$.)}
    \begin{adjustbox}{max width=1.0\textwidth}
    \setlength{\tabcolsep}{1.2mm}{
    \begin{tabular}{lllccc}
    \toprule
        \textbf{Problem} & \textbf{Dataset} &  \textbf{Description of Data Preparation} & \textbf{User Needs} & \textbf{Toxicity} $\downarrow$ & \textbf{Severe Toxicity }$\downarrow$\\
        \midrule
        \multirow{3}{*}{\parbox{2cm}{Conversational\\Information\\Seeking}}& Qulac \cite{qulac} & Created from the logs of search engine & Real & 0.052 & 0.004\\
        & Abg-CoQA \cite{abg-coqa} & Truncate conversations to induce ambiguity & Fabricated & 0.095 & 0.003\\
        & PACIFIC \cite{pacific} & Manually rewrite queries to induce ambiguity & Fabricated & 0.019 & 0.001 \\
        \midrule
        \multirow{3}{*}{\parbox{1cm}{Target-guided\\Dialogue}}&TGC \cite{acl19-tgc} & Rule-based keyword extractor to label targets  &  Fabricated & 0.197 & 0.020\\
        & TGConv \cite{coling22-topkg} & Randomly specify an easy target and a hard target & Fabricated & 0.202 & 0.012\\
        & DuRecDial \cite{acl20-durecdial} & Crowdworker annotations based on given user profiles & Fabricated & 0.118 & 0.007\\
        \midrule
        \multirow{3}{*}{\parbox{1cm}{Emotional\\Support\\Dialogue}}& HOPE \cite{hope} & Created from the transcriptions of counselling videos & Real & 0.151 & 0.007\\
        & MI \cite{mi} & Created from the transcriptions of counselling videos & Real & 0.122 & 0.005\\
        &ESConv \cite{esconv} & Crowdworker annotations based on given scenarios & Fabricated & 0.076 & 0.004\\
        \midrule
        \multirow{3}{*}{\parbox{1cm}{Negotiation\\Dialogue}}&CraigslistBargain \cite{emnlp18-negotiate} & Crowdworker annotations based on given bargaining targets & Fabricated & 0.160 & 0.011 \\
        & AntiScam \cite{aaai20-antiscam} & Crowdworker annotations based on given intents & Fabricated & 0.080 & 0.005\\
        & P4G \cite{acl19-p4g} & Crowdworker annotations with a pre-task survey as user profiles & Real & 0.048 & 0.002 \\
        \bottomrule
    \end{tabular}}
    \end{adjustbox}
    \label{tab:user_need}
    \vspace{-3mm}
\end{table*}

The potential of promoting genuine value to the system side ensures that the development of PCAs increasingly garners attention from both academic and industrial sectors. However, the societal acceptance of PCAs hinges crucially on meeting standards of \textsc{Adaptivity} and \textsc{Civility}. Consequently, it is of great importance to establish a well-defined objective for the foundation of PCAs.

\section{Data Preparation}\label{sec:data}

To gather conversational data, traditional methods often involve the recording and collection of raw dialogue samples, such as customer service logs, online forum threads, or video transcripts, capturing the data in its natural state. On the other hand, a significant portion of existing proactive dialogue datasets employs context-based data collection, either annotated by crowdworkers or generated by AI. Context-based data collection refers to the process of gathering dialogue data with underlying circumstances or background information that are pre-defined to direct the conversations. 

\vspace{-0.1cm}

\subsection{Issues on Current Data Preparation Schemes}\label{sec:data_analysis}
Apart from aiming at collecting data for the agent's \textsc{Intelligence}, we analyze the data preparation of existing proactive conversational datasets from the other two dimensions: \textsc{Adaptivity} and \textsc{Civility}. 

\vspace{-0.1cm}

\subsubsection{Fabricated User Needs}
From the perspective of \textsc{Adaptivity}, context-based data collection typically fabricates user needs for system-initiated behaviours to construct proactive conversation data. 
Upon training on the data with fabricated user needs for the agent's proactivity, it may result in inappropriate proactive behaviours regardless of real user needs, harming the adaptivity of PCAs.  
As listed in Table \ref{tab:user_need}, we analyze the data preparation process of several widely-studied datasets for various PCA tasks. 
For example, in conversational information seeking datasets, some ambiguous queries do not naturally occur, while the ambiguity is introduced by deliberately truncating conversations \cite{abg-coqa} or omitting information \cite{pacific}. 
Similar findings are drawn in a recent survey of asking clarification questions datasets \cite{acl23-clariq-survey}. 
In target-guided dialogue datasets \cite{acl19-tgc,coling22-topkg}, some target topics are assigned without considering actual user needs while just specifying some random words with unclear meanings, like "blue" or "tired". 
In emotional support dialogue datasets, 
dialogues can be produced by asking annotators to role-play as patients dealing with a specific imagined emotional problem \cite{esconv}, instead of the real user needs for counseling \cite{hope,mi}. 
Similar patterns are also found in negotiation dialogue \cite{emnlp18-negotiate,aaai20-antiscam} and target-guided dialogue datasets \cite{acl20-durecdial}, where annotators are instructed to play a pre-defined role. 
Overall, those datasets constructed by context-based crowdworker annotations or rule-based reconstructions are more likely to contain proactive dialogues with fabricated user needs. 

\vspace{-0.1cm}

\subsubsection{Ethical Concerns}\label{sec:toxicity}
Language toxicity has been an essential consideration in the perspective of \textsc{Civility} for human-centered PCAs. Following the toxicity assessment in previous studies \cite{acl23-augesc,extes}, we assess the toxicity of the utterances in the datasets listed in Table \ref{tab:user_need} by reporting the Toxicity and Severe Toxicity scores computed by \texttt{Perspective API} \cite{kdd22-perspectiveapi}. 
In general, crowdworker-annotated datasets exhibit safer conversations (lower toxicity scores) than those datasets collected from real-world conversations.  
For example, among three emotional support dialogue datasets, the toxicity degree of the two datasets collected from the raw transcriptions of counseling videos (\textit{i.e.}, HOPE \cite{hope} and MI \cite{mi}) are substantially higher than those of the crowdsourced dataset, ESConv \cite{esconv}.

\subsubsection{Issues on Different Types of Data Preparation Schemes}
Besides the above issues drawn from our analysis, we summarize the drawbacks of different types of data preparation schemes by further combining additional evidences from the literature as follows: \vspace{-0.1cm}
\begin{itemize}[leftmargin=*]
    \item \textbf{Real-world Collection}. (1) \textbf{Ethical Concerns}: Real-world data collection often involves sensitive personal information or potentially toxic content, raising significant privacy issues and necessitating toxicity assessment and detoxification. 
    (2) \textbf{Quality Variability}: Real-world conversations can vary greatly in quality, relevance, and clarity. 
    \item \textbf{Crowdworker Annotations}. (1) \textbf{Fabricated User Needs}: Crowdworkers are asked to perform conversations in a constrained setting, which could be different from how people with real user needs interact in natural conversations. (2) \textbf{Homogeneous Dialogue Patterns}: Crowdworkers generate dialogues specific to the provided context and are asked to generate dialogues of a certain style. Dialogues created in this way have a high degree of pattern overlap, either in lexical or logic \cite{naacl22-nlu++}.  
    \item \textbf{Generative AI Annotations}. (1) \textbf{Lack of Human Intuition}: AI-generated dialogue annotations by following human instructions \cite{aiia23-genai-annotation} might not capture the intricate contexts or subtleties of human conversation. For instance, while humans can grasp underlying emotions or subtext, AI might provide responses that feel shallow or off-mark. 
    (2) \textbf{Propagation of Biases in Dialogues}: If trained on a biased dialogue dataset, AI might reproduce and magnify these biases in the annotated dialogues \cite{emnlp20-bias}. This can lead to datasets that inadvertently favor or disfavor certain topics, demographics, etc.  
\end{itemize}

\subsection{New Perspectives on Data Preparation}
The dataset analysis in Section \ref{sec:data_analysis} reveals the limitations of existing data preparation schemes for proactive conversations. 
To mitigate these issues, we discuss two promising solutions as follows:

\subsubsection{Reflecting Real Human Needs}
A human-centered approach to data collection should emulate real-world scenarios to ensure the data genuinely represents actual human needs. 
A famous example is the data collection of Natural Questions \cite{tacl19-nq}, which consists of real anonymized, aggregated queries issued to the Google search engine. 
It revolutionizes the context-based QA data collection approaches \cite{emnlp16-squad} where annotators are asked to first read the passage containing the answer to generate the question. 
This is also valid under the context of proactive dialogue data collection. 
Collecting raw data from real-world conversations, like Qulac \cite{qulac}, HOPE \cite{hope}, and MI \cite{mi}, is the most straight-forward way to ensure real user needs. 
Similarly, 
\citet{sigir23-uneed-crs} construct a user needs-centric E-Commerce conversational recommendation dataset (U-NEED) from real-world E-Commerce pre-sales dialogues, where the target item is described by the real human users, instead of random target assignment. 
However, it can be difficult to obtain the desired resources and the collected real-world dialogues suffer from quality and ethical issues. 
In the case where annotators are necessary to construct a conversation dataset, the real user needs can be reflected by asking the annotators to just be themselves and collecting their own background information, such as using pre-task survey \cite{acl19-p4g}.

\subsubsection{Human-AI Collaborative Data Collection}\label{sec:human-ai-data}
To compensate for the limitations in both crowdworker and generative AI annotations, \citet{emnlp23-mathdial} pair human teachers with an LLM that simulates students and their errors for tutoring dialogue data collection. 
By integrating the nuanced understanding of human teachers with the scalable and controllable generation capabilities of LLMs, this approach can produce tutoring dialogue data with more diverse patterns and more educationally valuable intuition. 
In order to diversify the target accomplishment process with different user personalities for augmenting target-guided dialogue data, \citet{emnlp23-proactive-persona} create the \textsc{TopDial} dataset, which employs generative AI to simulate a variety of users by using Big-5 personality traits and user profiles.  
While improving generative AI annotations, an effective approach is to incorporate human knowledge for guiding the generative AI.  \citet{extes} leverage expert knowledge to design various emotional support strategies and collect real-world counseling cases for creating the emotional support dialogue dataset, named ExTES.  
Additionally, the issue of ethical concerns in real-world data collection can also be alleviated by the AI re-evaluation for conforming to social rule-of-thumbs \cite{emnlp22-prosocial}.

\begin{figure}[t]
\setlength{\abovecaptionskip}{-2pt}   
\setlength{\belowcaptionskip}{0pt}        
\centering      
\includegraphics[width=0.4\textwidth]{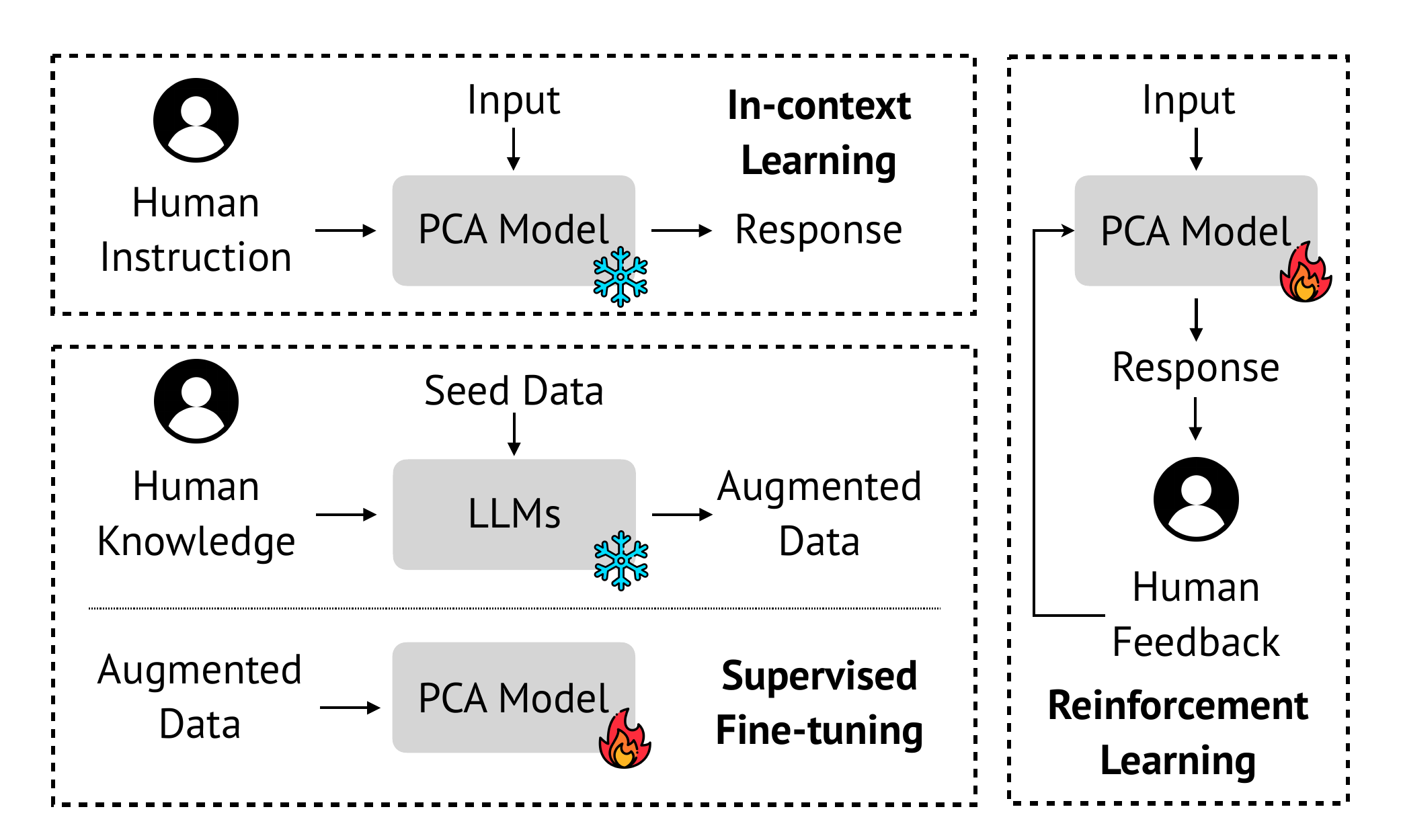}
\caption{Three types of human alignment approaches.}
\label{fig:method}
\vspace{-3mm}
\end{figure}

\section{Model Learning}\label{sec:model}
Most existing studies propose various advanced methods for building an intelligent proactive conversational agents that equip sophisticated planning capabilities. 
Despite the improved \textsc{Intelligence}, these PCAs are not always adept at interpreting a wide range of real-world situations or may produce responses that deviate from human expectations. 
While recently many efforts have been made in aligning language models with human values and expectations, namely \textit{Human Alignment} \cite{align-humanvalue,align-survey}, which is a valuable technique for integrating \textsc{Adaptivity} and \textsc{Civility} into the model learning of PCAs. 
As illustrated in Figure \ref{fig:method}, we discuss three main types of human alignment approaches for building human-centered PCAs, including In-context Learning (ICL), Supervised Fine-tuning (SFT), and Reinforcement Learning (RL). 
Furthermore, we prospect the remaining challenges and future research agenda accordingly.

\subsection{Prompting by Human Instructions}
ICL has emerged as a highly efficient learning paradigm in the era of LLMs, since LLMs possess substantial knowledge and exceptional instruction-following capabilities. 
\begin{itemize}[leftmargin=*]
    \item \textbf{\textsc{Intelligence}}. With the emerging  capabilities of LLMs \cite{tmlr-emerging}, plan generation by prompting LLMs is becoming the main-streamed paradigm for complex task solving. 
    Motivated by this, recent studies design various prompting schemes for instructing LLMs to conduct self-thinking of strategy planning, including Chain-of-Thought (CoT) \cite{emnlp23-procot,emnlp23-cuecot} and multi-agent debate \cite{acl23-askanexpert,negotiate-selfplay}. \\
    \textbf{Challenges}: Current prompt-based methods fail to do anticipation to optimize the long-term goal of the conversation. 
    \item \textbf{\textsc{Adaptivity}}. There is few work investigating the prompt-based approaches for taking account the \textsc{Adaptivity} of PCAs. 
    Notable observations are drawn from \citet{emnlp23-procot}, where the human-designed proactive CoT (ProCoT) prompting scheme mitigates the aggressive topic shift of LLM-based PCAs in target-guided dialogues. 
    Despite the lack of explicit designs for \textsc{Adaptivity}, the self-thinking instruction in ProCoT may enable the PCA to capture nuances in conversations, such as user satisfaction.\\ 
    \textbf{Challenges}: The underlying reasons for the enhanced \textsc{Adaptivity} remain unclear and the improvement is still far from satisfactory where LLMs with ProCoT prompting still tend to make more aggressive topic transitions than desired. 
    \item \textbf{\textsc{Civility}}. By integrating the alignment goals directly into the prompts, ICL can guide and regulate the responses of PCAs, ensuring they align more closely with desired outcomes and guidelines. For example, \citet{acl23-mi-prompt} empirically show that PCAs with manually designed mixed-initiative strategy prompts become more honest and thoughtful (higher \textsc{Civility}) in emotional support and persuasion dialogues. \\
    \textbf{Challenges}: The manually designed prompts for each type of strategy lack transferability to unseen scenarios and are limited to specific abilities, such as Manners and Moral Integrity in \citet{acl23-mi-prompt}, while neglecting other abilities of \textsc{Civility}. 
\end{itemize}
\vspace{-0.1cm}

\subsection{Data Augmentation with Human Knowledge}
The utilization of LLMs for data augmentation in conversational systems has gained substantial attention for offering promising avenues to improve response quality and dialogue performance. 
\begin{itemize}[leftmargin=*]
    \item \textbf{\textsc{Intelligence}}. There are two typical paradigms: (1) Self-chat distillation methods \cite{acl23-augesc,extes} directly distill the conversational intelligence from LLMs by prompting LLMs to complete the conversations with specific human instructions. 
    (2) Role-play simulation methods \cite{emnlp23-mathdial,emnlp23-proactive-persona} employ LLMs to simulate a specific role in the conversation to communicate with humans or other role-playing agents for collecting  conversation data. \\
    \textbf{Challenges}: Despite the remarkable quality of LLM-augmented dialogue data, this type of data inevitably inherits the limitation of LLMs in handling proactive dialogues, such as limited abilities to make strategic decisions and plans for long-term goals. 
    \item \textbf{\textsc{Adaptivity}}. The practice of SFT using limited annotated datasets may result in a lack of generality in PCAs, particularly when encountering diverse user personalities or unforeseen scenarios.  
    To address this, recent research \cite{emnlp23-proactive-persona,extes} has increasingly focused on incorporating human knowledge into LLMs to enhance the \textsc{Adaptivity} of these systems, allowing them to better handle a wider range of conversations and user needs. 
    \\
    \textbf{Challenges}: The generality of the augmented data is still limited by the available human knowledge. 
    \item \textbf{\textsc{Civility}}. To enable PCAs appropriately respond to unsafe or unethical user utterances, \citet{emnlp22-prosocial} augment morality-related dialogue datasets with social rule-of-thumb knowledge from human annotators. 
    After being fine-tuned on the augmented data, PCAs can lead the conversation in a prosocial manner. \\
    \textbf{Challenges}: Due to the absence of negative feedback from humans, the fine-tuned model only knows what should do but may fail to prevent from what should not do for achieving \textsc{Civility}. 
\end{itemize}
\vspace{-0.1cm}

\begin{table*}[]
\setlength{\abovecaptionskip}{2pt}   
\setlength{\belowcaptionskip}{2pt}
    \centering
    \caption{Experimental results on the ESConv dataset \cite{esconv}. (The lower the better $\downarrow$. The higher the better $\uparrow$.)}
    \begin{adjustbox}{max width=1.0\textwidth}
    \setlength{\tabcolsep}{1.2mm}{
    \begin{tabular}{llcccccccccc}
    \toprule
    \multirow{2}{*}{\textbf{Type}}  &\multirow{2}{*}{\textbf{Method}}  & \multicolumn{2}{c}{\textsc{\textbf{Intelligence}}} & \multicolumn{3}{c}{\textsc{\textbf{Adaptivity}}} & \multicolumn{5}{c}{\textsc{\textbf{Civility}}} \\
    \cmidrule(lr){3-4}\cmidrule(lr){5-7}\cmidrule(lr){8-12}
    && Succ. Rate $\uparrow$ & Avg. Turn $\downarrow$ & Smoothness $\uparrow$ & Satisfaction $\uparrow$  & ECE $\downarrow$ & Toxicity $\downarrow$ & Identity Attack $\downarrow$ & Threat $\downarrow$ & Insult $\downarrow$ & Relaxation $\uparrow$ \\
    \midrule
    - & ChatGPT \cite{chatgpt} & 0.7692 & 5.10 & 0.3933 & 4.29 & 0.4631 &0.0591 & 0.0019 & 0.0105 & 0.0261 & 0.3773 \\
    \midrule
    \multirow{2}{*}{ICL} & Ask-an-Expert \cite{acl23-askanexpert} & 0.8000 & 4.76 & 0.3346 & 4.16 & 0.3814 & 0.0633 & 0.0082 & 0.0089 & 0.0284 & 0.3958\\
    & ProCoT \cite{emnlp23-procot} & 0.7769 & 4.83 & 0.3704 & 4.26 & \textbf{0.3199} & 0.0586 & \textbf{0.0061} & \textbf{0.0080} & 0.0265 & 0.3525 \\
    \midrule
    \multirow{2}{*}{SFT} & AugESC \cite{acl23-augesc} & 0.7445 & 5.43 & 0.4181 & 3.80 & 0.3856 & 0.0605 & 0.0086 & 0.0184 & 0.0254 & 0.3482 \\
    & ExTES \cite{extes} & 0.7954 & 4.67 & 0.4437 & \textbf{4.35} & 0.3321 & \textbf{0.0526} & 0.0071 & 0.0082 & \textbf{0.0071} & \textbf{0.4110} \\
    \midrule
    \multirow{2}{*}{RL} & RLHF \cite{instructgpt} & 0.8592 & 4.51 & 0.4398 & 3.92 & 0.4053 & 0.0629 & 0.0100 & 0.0245 & 0.0273 & 0.3851 \\
    & Aligned$_\text{d-PM}$ \cite{nips23-align} & \textbf{0.8785} & \textbf{4.46} & \textbf{0.4525} & 4.09 & 0.3816 & 0.0554 & 0.0065 & 0.0080 & 0.0275 & 0.4092\\
    \bottomrule
    \end{tabular}}
    \end{adjustbox}
    \label{tab:exp}
    \vspace{-4mm}
\end{table*}

\subsection{Learning from Human Feedback}
Reinforcement learning from Human Feedback (RLHF) \cite{instructgpt} is designed to align the language model with humans from human preference signals under the RL framework.
\begin{itemize}[leftmargin=*]
    \item \textbf{\textsc{Intelligence}}. Due to the high cost of human feedback on the intelligent completion for long-term goals, researchers \cite{ppdpp,gdpzero}  simulate the goal-oriented human feedback by using generative AI for enhancing the planning and anticipation abilities of PCAs. \\
    \textbf{Challenges}: RLHF assumes that human advances in their abilities for aligning the model behaviours. However, considering a future PCA model may become much more intelligent than humans, humans will no longer be reliable to supervise the model in those complex tasks that we don't understand. This situation raises the needs for exploring Superalignment  approaches \cite{superalignment} that ensure superintelligent PCAs reliably follow human intents. 
    \item \textbf{\textsc{Adaptivity}}. Most RLHF approaches generally use majority voting or averaging to combine inconsistent preferences into a unified one. However, this process represents only a narrow segment of individuals, failing to effectively reveal the full scope of human preferences universally. To remedy this, \citet{nips23-align} propose to account for the distribution of disagreements among human preferences. \citet{trip-non-collab} designs role-playing user simulators with various personality to interact with PCAs for enhancing the diversity of human feedback. \\
    \textbf{Challenges}: The cost of collecting diverse human feedback is high and the alignment relies heavily on the quality of human preference feedback. Finding a balance between minimizing human cost while still maintaining high-quality alignment is a key challenge in the development of human-aligned \textsc{Adaptivity}.
    \item \textbf{\textsc{Civility}}. Besides the general safety alignment in most RLHF approaches, \citet{opo} further propose to employ an external memory to store established rules for alignment with diverse and customized human values, including legal and moral rules. \\
    \textbf{Challenges}: As for PCAs, the measurement of feedback quality is supposed to be multifaceted, involving the three key dimensions. Therefore, the widely-used binary feedback becomes indistinguishable in quality, while it is critical to investigate a more comprehensive form of human feedback collection for PCAs.
\end{itemize}
\vspace{-0.1cm}

\section{Evaluation}\label{sec:eval}

An ideal human-centered proactive conversational agent should fulfill certain criteria across three key dimensions. 
However, existing evaluations of PCAs mainly focus on the \textsc{Intelligence} perspective, using metrics like Goal Completion and Response Quality. 
According to various goals, the evaluation of goal completion can be referred to 
the achievement of the target topic~\cite{coling22-topkg,sigir22-proactive}, the clarification-based information seeking \cite{wsdm22-eval-mix}, the negotiation gain \cite{iclr21-negotiate,acl19-p4g}, etc. 
Meanwhile, the quality of responses is typically judged by human annotators, involving factors like fluency, informativeness, etc.
However, the other two aspects are often overlooked.

Due to the lack of existing metrics for assessing the representative abilities of \textsc{Adaptivity} and \textsc{Civility}, we propose a preliminary multidimensional evaluation framework by adopting some alternative evaluation protocols to evaluate these abilities for PCAs. 
Specifically, we take a widely-studied PCA task as a case study, namely Emotional Support Dialogues introduced in Section \ref{sec:task}, and conduct the multidimensional evaluation on different human-centered model learning techniques discussed in Section \ref{sec:model}.

\subsection{Proposed Evaluation Framework}\label{sec:eval_metric}
Besides \textsc{Intelligence} which has been extensively evaluated in the literature, including reference-based response quality metrics (\textit{e.g.,} BLEU and ROUGE) and goal completion metrics (\textit{e.g.,} Success Rate and Average Turn \cite{ppdpp}), we first introduce some alternative evaluation protocols for the evaluation of  \textsc{Adaptivity} and \textsc{Civility}. 
\subsubsection{Evaluation Protocols for \textsc{Adaptivity}}
There are three representative abilities of \textsc{Adaptivity}:
\begin{itemize}[leftmargin=*]
    \item \textbf{Patience} refers to the ability to adaptively manage the pace of taking initiative, which is intrinsically linked to the smoothness of the conversation. Following some existing studies of PCAs \cite{coling22-topkg,emnlp23-procot}, we measure the smoothness by the contextual semantic similarity between the last utterance and the generated response.   
    \item \textbf{Timing Sensitivity} refers to the ability to take initiative accounting for real-time user needs and status, which 
    can be alternatively evaluated by the user satisfaction at each conversation turn. There are different approaches to measure user satisfaction, such as topic-based scoring \cite{sigir22-proactive}, data-driven estimation \cite{www22-use,acl23-use}, and LLM-based prediction \cite{cikm23-user-sim} (adopted). 
    \item \textbf{Self-awareness} refers to the ability to recognize its own limitations. Inspired by uncertainty calibration studies \cite{acl23-calibration,emnlp23-calibration}, we adopt Expected Calibration Error (ECE) to measures how well an agent’s confidence match the observed accuracy. 
    Specifically, we regard the success rate as the accuracy and the probability of taking initiative behaviour as the confidence of PCAs. 
\end{itemize}
\vspace{-0.1cm}

\subsubsection{Evaluation Protocols for \textsc{Civility}}
There are five representative abilities of \textsc{Civility}, including Boundary Respect, Moral Integrity, Trust and Safety, Manners, and Emotional Intelligence. Similar to the analysis in Section \ref{sec:toxicity}, we also adopt the \texttt{Perspective API} \cite{kdd22-perspectiveapi} for automatically scoring the first four abilities based on the corresponding attributes of Identity Attack, Toxicity, Threat, and Insult. 
As for the last ability, \textit{i.e.}, Emotional Intelligence, we adopt Emotional Intensity Relaxation \cite{acl23-kemi} for evaluation.

\begin{figure}[t]
\setlength{\abovecaptionskip}{0pt}   
\setlength{\belowcaptionskip}{0pt}          
\centering      
\includegraphics[width=0.43\textwidth]{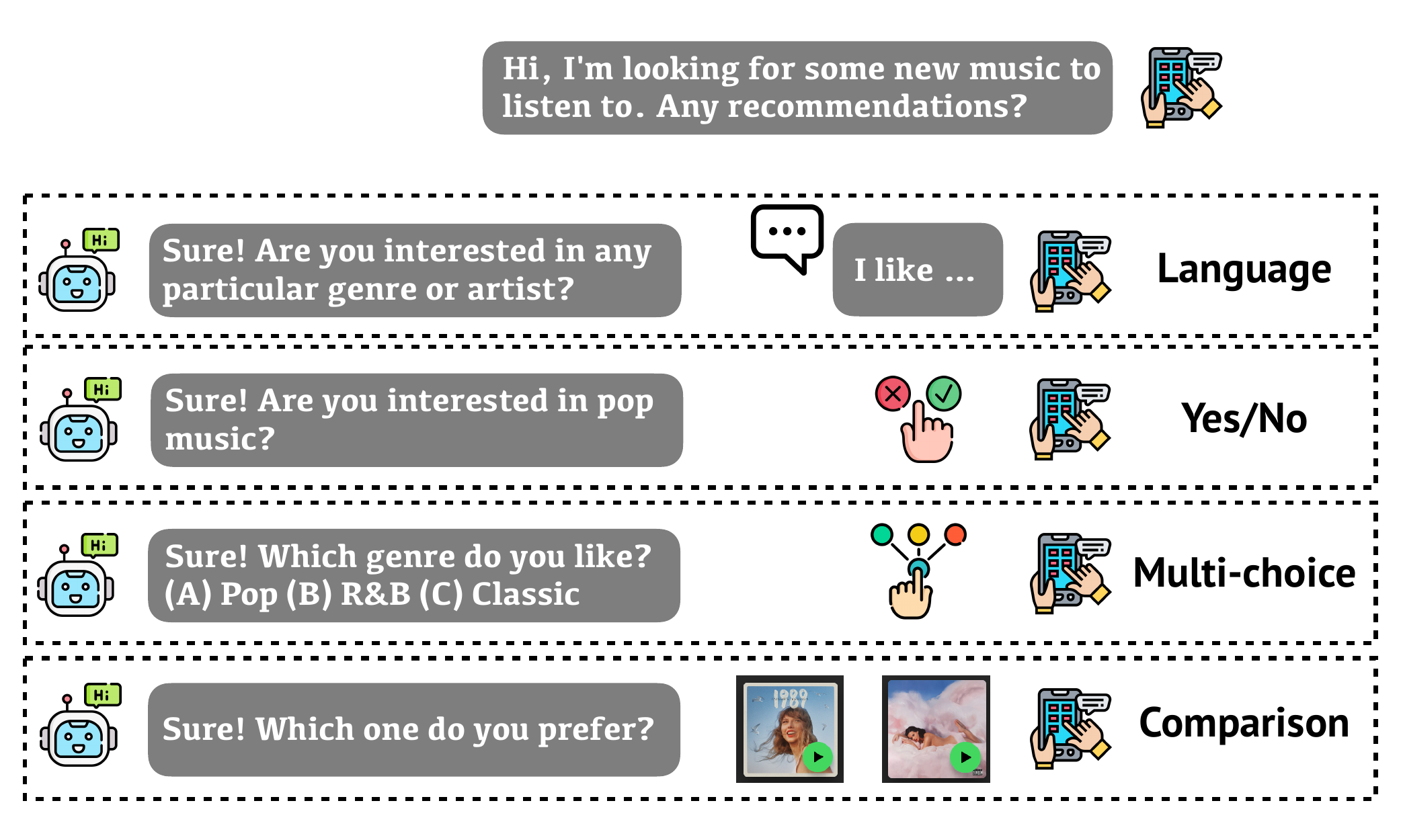}
\caption{Different user interface designs for user preference elicitation in conversational recommender systems.}
\label{fig:interface}
\vspace{-4mm}
\end{figure}

\subsection{Empirical Analysis} 
We adopt a widely-studied emotional support dialogue dataset, ESConv \cite{esconv}, as the testbed for analysis. 
According to the three main types of human alignment approaches in Section \ref{sec:model}, we adopt corresponding methods for evaluation, including two prompt-based methods (Ask-an-Expert \cite{acl23-askanexpert} and ProCoT \cite{emnlp23-procot}), two supervised fine-tuning methods (AugESC \cite{acl23-augesc} and ExTES \cite{extes}), and two RL-based methods (RLHF \cite{instructgpt} and Aligned$_\text{d-PM}$ \cite{nips23-align}). 
Almost all the existing studies only evaluate the PCA from the dimension of \textsc{Intelligence}, while in our analysis, we further include the aforementioned evaluation protocols for assessing the agents' abilities from the dimensions of \textsc{Adaptivity} and \textsc{Civility}. 
Experimental results are summarized in Table \ref{tab:exp}. 
There are two notable observations from the results: 
\begin{itemize}[leftmargin=*]
    \item \textbf{Methods perform diversely in terms of metrics of three dimensions.} For example, although Aligned$_\text{d-PM}$ achieves the best performance in \textsc{Intelligence}, \textit{i.e.}, it still under-performs in some metrics in the other two dimensions. 
    \item \textbf{It attaches great importance in involving the three dimensions into model learning.} Compared with the standard RLHF, Aligned$_\text{d-PM}$ actually improves the performance in the dimension \textsc{Adaptivity} by taking into account the diversified user preferences. Similarly, ExTES performs better than AugESC in \textsc{Adaptivity} by using human knowledge to design various strategies and collect real-world counseling cases. 
\end{itemize}
\vspace{-0.1cm}

\subsection{Prospects on Multidimensional Evaluation}
As discussed in the empirical analysis, higher scores in terms of \textsc{Intelligence}-related metrics is not necessarily accompanied with better performance in \textsc{Adaptivity} and \textsc{Civility}. This helps us understand the limitations of current evaluation methods and sheds light on future opportunities. 

\subsubsection{Robust Evaluation Protocols} Due to the subjective nature and long-term impact for the evaluation of \textsc{Adaptivity} and \textsc{Civility}, human interactive judgments remain the most effective method for assessing these two human-centered dimensions. However, human judgements are challenging to scale and lack standardization for comparisons. 
For example, \citet{huang2024concept} integrates both system- and user-centric factors into the evaluation of conversational recommender systems. 
Our proposed taxonomy offers a solid foundation for developing suitable automatic evaluation metrics. While we explore various alternative metrics in Section \ref{sec:eval_metric}, there is a clear need for more reliable and robust automatic metrics that seamlessly reflect the attributes of human-like conversational agents. 

\subsubsection{Customized Evaluation Framework} Different types of PCAs will need different evaluation gold standard, rather than solely evaluating on the \textsc{Intelligence} dimension or requiring high proficiency in all three dimensions. 
For example, for some \texttt{Listener}-type PCAs, like social chatbots, even \textsc{Intelligence} is not a necessary evaluation standard as humans may only need a listener who can just listen to their stories with limited initiative and planning capabilities. 
In certain emergency scenarios, the importance of respecting boundaries may diminish, as indicated by the user study in \citet{cui22-voice-assistant} which reveals that concerns about privacy can be overshadowed by the  physical safety. 
Therefore, it is crucial to customize the human-centered evaluation framework of PCAs for different social contexts or applied scenarios.

\section{System Deployment}
\label{deploy}
From the perspective of ubiquitous computing \cite{weiser1991computer}, the most profound human-centered proactive conversational agents are those that seamlessly integrate into daily life, becoming a natural part of it. 
To achieve this, the design of human-centered system deployment should focus more on human behaviour patterns rather than just extending technology functionalities. 

\subsection{Human-centered Designs of User Interface}

While conversational agents often prioritize language as the main interface, this design can be imprecise for tasks needing specific user inputs. 
Language-based interactions can be challenging to control for desired outcomes and might be unsuitable for precision-required tasks. 
Additionally, these interactions can sometimes be inconvenient and inefficient for quick or straightforward tasks, where simpler interfaces might be more effective. Balancing language interfaces with other interaction modalities can enhance both user experience and task efficiency.
In the realm of PCAs, it's essential to carefully design user interfaces that are minimally intrusive for system-initiated interactions. This consideration ensures a balance between functionality and user comfort.

Take conversational recommender systems as an example. 
The system-initiated interactions in conversational recommender systems typically refer to user preference elicitation, which aims to explicitly acquire user preference rather than solely inferring users' implicit preference from the conversation history. 
As shown in Figure \ref{fig:interface}, the most popular paradigm is called "System Ask, User Respond" \cite{cikm18-crs}, where the PCA uses \textbf{language} as the interface to ask eliciting questions for collecting users' preference descriptions. 
However, the language interface of PCAs faces several challenges in conversational recommender systems: 1) understanding user preferences from natural language itself is a challenging problem, and 2) users may feel uncomfortable to frequently provide their personal information to the technology company. 
In recent years, many other types of user preference elicitation interfaces have been studied, including asking \textbf{Yes/No} questions or \textbf{multiple-choice} questions \cite{www22-mcq-crs,wsdm20-ear} for identifying users' preference towards specific item features, and presenting \textbf{comparisons} \cite{sigir21-comparative-crs} for obtaining relative preference feedback. 
In this manner, human users can seamlessly and precisely convey their preference by tapping on certain icons on their devices to interact with PCAs. 

\subsection{New Emphasis on Trust and Reliance}

\begin{figure}[t]
\setlength{\abovecaptionskip}{0pt}   
\setlength{\belowcaptionskip}{2pt}         
\centering      
\includegraphics[width=0.35\textwidth]{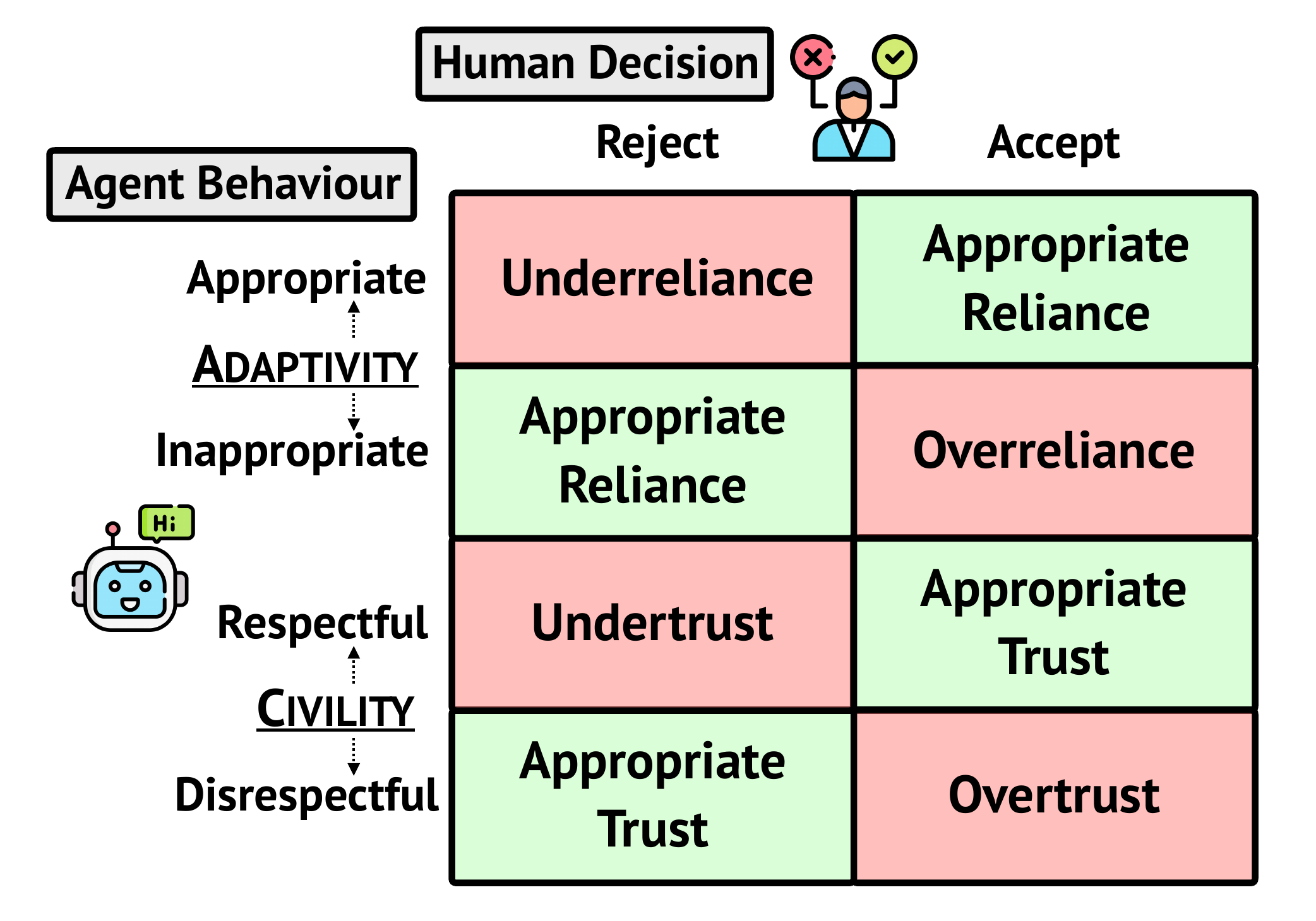}
\caption{A simplified diagram outlining the different ways people rely on and trust proactive conversational agents.}
\label{fig:trust}
\vspace{-4mm}
\end{figure}

\begin{figure}[t]
\setlength{\abovecaptionskip}{2pt}   
\setlength{\belowcaptionskip}{2pt}         
\centering      
\includegraphics[width=0.46\textwidth]{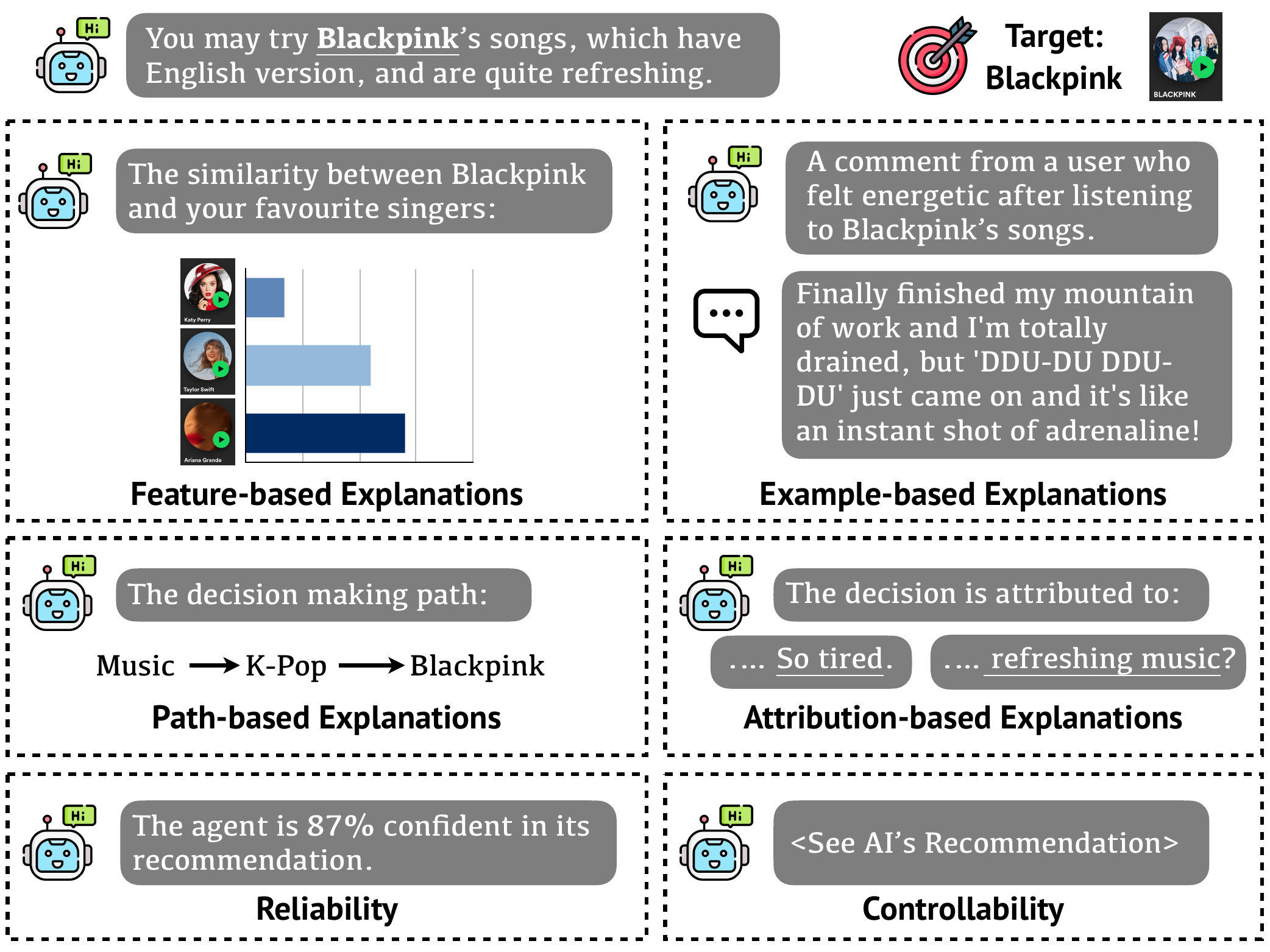}
\caption{Example UI designs regarding appropriate trust and reliance for target-guided conversational recommendation.}
\label{fig:ui}
\vspace{-3mm}
\end{figure}

When interacting with human-centered proactive conversational agents, users should feel comfortable using and depending on the system for proactively achieving their goals. 
Inspired by HCI studies \cite{cscw21-trust,cscw23-explanability}, human-centered PCAs should exhibit appropriate trust and reliance from human users. 
As illustrated in Figure \ref{fig:trust}, when provided with a system-initiated behaviour from a proactive conversational agent, the human user has the choice to either accept or reject the interaction. 
The agent may exhibit appropriate or inappropriate behaviours in terms of their \textsc{Adaptivity}, and also respectful or disrespectful behaviours in terms of their \textsc{Civility}. 
\begin{itemize}[leftmargin=*]
    \item \textbf{Appropriate Reliance}: the human accepts appropriate agent behaviours or corrects inappropriate agent behaviours. 
    \item \textbf{Underreliance}: fails to accept appropriate agent behaviours. 
    \item \textbf{Overreliance}: fails to correct inappropriate agent behaviours.
    \item \textbf{Appropriate Trust}: accepts respectful agent behaviours or rejects disrespectful agent behaviours. 
    \item \textbf{Undertrust}: refuses to accept respectful agent behaviours. 
    \item \textbf{Overtrust}: fails to reject disrespectful agent behaviours.
\end{itemize}
We advocate human-agent interactions with appropriate reliance and trust, since underreliance and undertrust  can impede the goal achievement while overreliance and overtrust may pose risks to human users. 
As presented in Figure \ref{fig:ui}, we discuss several approaches from the HCI perspective to address this issue.

\subsubsection{Explanability}
Explanations help bridge the gap between complex AI decision-making processes and human understanding. 
Extensive HCI studies validate that presenting explanations can reduce users' overreliance \cite{cscw23-explanability,cscw23-explantion-example} and overtrust \cite{icwsm21-overtrust-exp} on the prediction of AI systems.  
In the context of human-centered PCAs, we introduce four types of explanations that can be presented to human users. 1) Feature-based explanations show the contribution of different features to the decision making of PCAs. 2) Example-based explanations present either representative prototypes of the decision that the PCA makes for the given instance or examples that are similar to the given instance along with the PCA's decision. 3) Path-based explanations present the decision path made by the PCA between the initial state and the current state. 4) Attribution-based explanations attributes the output decision to specific parts or components of the input data or external knowledge. 

\subsubsection{Reliability}
Another key finding in HCI studies \cite{hf09-reliability,chi19-accuracy,cscw21-trust} is the importance of "reliability disclosure" in shaping users' trust and reliance on system feedback. This concept involves explicitly informing users about the estimated reliability or uncertainty of the system's feedback. 
For example, human users were shown the system’s confidence along with the decision, like "\textit{The agent is 87\% confident in its suggestion}". 
Meanwhile, how well the system can estimate its confidence is also a challenging problem, known as self-calibration. 
This involves a range of methods to enhance the system's ability to assess and communicate its own confidence \cite{tacl22-calibration-conv,emnlp23-calibration}. 
When users are aware of an agent's confidence, as assessed by the agent itself, it significantly impacts their trust and reliance on the conversational agent.

\subsubsection{Controllability}
In conventional designs, proactive conversational agents often hold complete autonomy in deciding when to initiate interactions, leaving users with little choice but to engage with these system-initiated behaviors. 
However, some user studies in HCI literature \cite{cscw21-trust,chi22-trust-crs} suggests the importance of empowering users with control over these interactions. Allowing users the flexibility to determine the necessity of system-initiated behaviors can significantly address issues of underreliance and undertrust. 
For example, the system-initiated behaviours, such as asking clarification questions or providing suggestions, are on demand and not presented to the users by default, while users could choose to see this content by clicking on a button, like "\textit{See AI's suggestion}". 

\section{Conclusions}
This perspectives paper investigated proactive conversational agents from the human-centered perspective. 
We first proposed a new taxonomy concerning three key dimensions of human-centered PCAs, including \textsc{Intelligence}, \textsc{Adaptivity}, and \textsc{Civility}. 
According to this taxonomy, we re-interpreted existing literature on PCAs upon the five stages of PCA system construction (\textit{i.e.}, Task Formulation, Data Preparation, Model Learning, Evaluation, and System Deployment). 
In the light of the limitations, we envisioned future research agenda and prospects for achieving human-centered PCAs. 
Meanwhile, PCAs are advancing towards the realm of superintelligent AI, where maintaining a human-centered system is crucial to ensure these superintelligent AIs continue to serve human's interests. 

\section*{Acknowledgement}
This research was supported by NSF grant III-2336768 and NExT Research Center. This research was also supported by the Singapore Ministry of Education (MOE) Academic Research Fund (AcRF) Tier 1 grant (Proposal ID: 23-SIS-SMU-010). 

\bibliographystyle{ACM-Reference-Format}
\bibliography{sample-base}

\appendix

\end{document}